\documentclass{elsart}
\usepackage[]{epsfig}
\begin{document}

\begin{frontmatter}

\title{Why the length of a quantum string cannot be Lorentz contracted.}

\author{Antonio Aurilia\thanksref{s}}
\thanks[s]{e-mail address: aaurilia@csupomona.edu }
\address{Department of Physics and Astronomy,
  California State Polytechnic University, Pomona, USA }

\author{Euro Spallucci\thanksref{infn}}
\thanks[infn]{e-mail address: spallucci@ts.infn.it }
\address{Dipartimento di Fisica, Sezione Teorica, Universit\`a di Trieste, 
and INFN, Sezione di Trieste, Italy}

\begin{abstract}
We propose a quantum gravity-extended form of the classical length contraction
law obtained in Special Relativity. More specifically, the framework of our discussion is the 
\emph{UV self-complete theory of quantum gravity.} Against this background, we show how our results are consistent with, 
i) the generalized form of the Uncertainty Principle (GUP), ii) the so called \emph{hoop-conjecture} which we interpret, 
presently, as the saturation of a Lorentz boost by the formation of a black hole in a two-body scattering, 
and iii) the intriguing notion of ``classicalization'' of trans-Planckian physics. Pushing these ideas to their logical
conclusion, we argue that there is a physical limit to the Lorentz contraction rule in the form of 
some \emph{minimal universal length} determined by quantum gravity, say the Planck Length, or any of its current 
embodiments such as the string length, or the $TeV$ quantum gravity length scale.
In the latter case, we determine the \emph{critical boost} that separates the ordinary ``particle phase,'' 
characterized by the Compton wavelength, from the ``black hole phase'', characterized by the effective Schwarzschild radius
of the colliding system. 
Finally, with the ``classicalization'' of quantum gravity in mind, we comment on the remarkable \emph{identity,} to our 
knowledge never noticed before, between three seemingly independent \emph{universal} quantities, namely, a) the ``string tension'', 
b) the ``~linear energy density,~'' or \emph{tension} that exists at the core of all Schwarzschild black holes, and c) 
the ``superforce'' i.e., the Planckian limit of the static electro-gravitational force and, presumably, the unification 
point of all fundamental forces. 
\end{abstract}
\end{frontmatter} 

\section{Introduction and background}
High energy particle physics is  based on the notion
that smaller and smaller distance scales can be investigated by increasing
the energy of the probe particle. Elementary projectiles colliding with a target 
 can resolve distances comparable with their quantum mechanical wavelength.
 The more is the energy, the shorter is the wavelength in agreement with the relativistic rule of length contraction. 
Quantum mechanics  and  Special Relativity work together to open a window on the microscopic world.\\
 This simple picture becomes less clear when we imagine to approach the
 Planck scale of distance, or energy, and consider the concomitant \emph{quantum gravity effects}\cite{uno}
This problem has long been ignored on the basis that the Planck energy, roughly $10^{19} GeV $, is so huge that no particle 
accelerator will ever be able to approach it.  \\
However, the picture is completely different when we consider the
string-inspired unified models with large extra-dimensions, where the
unification scale can be as low as some $TeV$.  In this kind of scenario,
quantum gravity effects, including micro black hole production in partonic
hard scattering, have been suggested to occur near the LHC peak energy,
i.e., $14\, TeV$ 
\cite{Giddings:2001bu,Cavaglia:2002si,Rizzo:2006zb,Casanova:2005id,Casadio:2008qy,Gingrich:2010ed,Nicolini:2011nz,Bleicher:2010qr},
\cite{Spallucci:2012xi,Mureika:2011hg}.  In  this new physics the distinction between ``point-like''
elementary particles and ``extended'' quantum gravity excitations, whatever
they are, i.e., black holes, $D$-branes, string balls, etc., turns out to be fuzzy, so that standard notions, such as the 
Lorentz-Fitzgerald length contraction, require a substantial revision, at least insofar as its domain of validity is concerned. 
For instance, a fundamental, quantum string
is, presumably, the smallest object in Nature with a linear size given by $l_s =\sqrt{\alpha^\prime}$. Thus, in this 
\emph{string perspective,} no distance
shorter than $l_s$ can be given a physical meaning. Furthermore, by supplying more and more energy,
higher and higher vibration modes are excited making the string longer and longer, in conflict with the length-contraction 
rule, but not unlike the increasing size of a Schwarzschild black hole. To our mind, this signals the end of validity of 
special relativity and the onset of gravitational effects.\\
How can we account for that?\\
Before trying to answer this question, it is useful to recall the derivation of the fundamental units that define the domain 
of quantum gravity, as the answer to our question lies in the very definition of those units.\\   

The appropriate standards of length, mass and time were originally introduced by Max Planck on a purely dimensional
basis by combining the speed of light $c$, the Gravitational coupling constant $G_\ast$
\footnote{ $G_\ast$ can be either the Newton constant, or the higher-dimension gravitational
coupling of $TeV$ quantum gravity.} and the Planck constant $\hbar$. In other words, Planck recognized that it is possible 
to combine Special Relativity, Quantum Mechanics and Gravity in the following dimensional package,

\begin{equation}
L_P \propto \sqrt{\frac{\hbar G_\ast}{c^3}}\ ,\quad
T_P\propto \sqrt{\frac{\hbar G_\ast}{c^5}}\ ,\quad
M_P\propto \sqrt{\frac{\hbar c}{G_\ast}} \label{pdimens}
\end{equation}

Clearly, this dimensional approach defines the Planck units up to
a numerical factor providing only  an ``orders of magnitude'' estimate.
In the old days, the Planck world was envisaged as the arena of violent
quantum gravity fluctuations disrupting the very fabric of space and time
\cite{Wheeler:1998vs}.
Eventually, the notion of ``space-time foam'' evolved into a ``Planckian phase'' with a different
description according to String/M-Theory, Loop Quantum gravity, Non-Commutative
geometry, Fractal space-time, etc.\\
In order to determine the numerical constants in (\ref{pdimens}) some extra argument is due.\\
For instance, one may declare that the Planck Mass is defined  by the equality
between the quantum mechanical wavelength of a particle and its gravitational
critical radius:

\begin{equation}
\frac{\hbar}{M_\ast c} = \frac{2 M_\ast G_\ast}{c^2}\label{intersect}
\end{equation}

Thus,

\begin{equation}
L_\ast = \sqrt{\frac{2\hbar G_\ast}{c^3}}\label{lplanck}\ ,\quad
M_\ast = \sqrt{\frac{\hbar c}{2 G_\ast}} \label{lmplanck}
\end{equation}

An alternative, but consistent, definition of (\ref{lmplanck}), which to our knowledge has never been noted before,
is the following: $L_\ast$ is the \emph{geometric mean} of the quantum mechanical wavelength
  $\lambda_C=\hbar/mc $ of the particle and its critical gravitational radius $R_s=2m G_\ast/c^2$

\begin{equation}
L_\ast \equiv \sqrt{\lambda_C R_s}=\sqrt{\frac{2\hbar G_N}{c^3}}
\end{equation}

Further insight into the physical meaning of $L_\ast $  can be obtained 
from the Generalized Uncertainty Principle (GUP)\cite{Kempf:1994su,Kempf:1996nk,Witten:1996,Adler:1999bu,Adler:2001vs}, 
where $L_\ast $ is often identified with the string length, i.e., $L_\ast= \sqrt{\alpha^\prime} $ 

\begin{equation}
\Delta x \ge \frac{\hbar}{\Delta p } + \frac{L_\ast^2}{4}
\frac{\Delta p }{\hbar}\label{gup}
\end{equation}

By minimizing the uncertainties, one finds

\begin{equation}
\Delta p_\ast =\frac{2\hbar}{L_\ast} \ ,\qquad\Delta x_\ast=L_\ast \label{deltax}
 \end{equation}

From  equation  (\ref{deltax}) we see that $L_\ast$ represents the minimal uncertainty in the particle/string
localization. From this point of view, $L_\ast$ is the \emph{minimal length} which is physically meaningful since, for a 
shorter one, the uncertainty is larger than the length itself. In contrast to this, it seems worth observing that the 
Planck mass is neither an absolute minimum nor an absolute maximum. It is, rather, an \emph{extremal value,} or turning point, 
in the sense that, as implied by its definition (\ref{intersect}), it represents the largest mass that an elementary 
particle may possess, or the smallest mass attributable to a micro-black hole. Interestingly enough, we shall argue in the 
following section, as well as in the last section of this article that 
there exists in nature a universal, unsurpassable \emph{linear energy density,} or tension, that lies at the core 
of every black hole, regardless of its mass or size.\\

\section{Critical boost and minimal length}

We have remarked earlier that the existence of a ``\emph{quantum of length}'' \cite{Garay:1994en,Sprenger:2012uc} 
is in conflict with 
the conventional rule of ``~length contraction~'' derived in special relativity. In a nutshell, the quantum of length 
$L_\ast $ is a new universal constant on the same footing as $c$ and $\hbar$, and as such it must be observer
independent. It follows that $L_\ast$ must act as an unbreakable barrier to the Lorentz-Fitzgerald contraction.\\
We propose to get around this problem by redefining  the Lorentz-Fitzgerald contraction law
 in the presence of a short-distance Planck barrier.  This is the crux of the following discussion.\\
In Special Relativity a rod of length $L_0$ in its rest frame is seen to be contracted in the direction of motion
according to the rule:

\begin{equation}
 L\left(\, \beta\,\right) = L_0\, \sqrt{1-\beta^2}\ , \qquad \beta\equiv v/c \label{contract}
\end{equation} 

An immediate consequence of (\ref{contract}) is that $L$ can contract to an arbitrarily small length as $\beta\to 1$. 
There are, however, at least two types of objections to this conclusion that require a redefinition of the contraction rule:
\begin{itemize}
 \item ``Quantum'' objection, or, the absence of $\hbar$: even though equation (\ref{contract}) is routinely applied to 
the world of particle physics, it was conceived with \emph{macroscopic}, i.e. ``classical'', objects in mind.
Stated otherwise, the quantum of action $\hbar$ seemingly has no effect in the length-contraction rule, but we expect this 
to change in the ultra-relativistic regime when one approaches distances of the order of the Planck length.
\item ``Gravitational'' objection, or, the absence of $G_N$ :  equation (\ref{contract}) refers to ``~abstract~'' lengths  
ignoring the fact that any physical object produces its own gravitational field, and thus introduces a ``critical'' 
gravitational length scale, that is, 
the Schwarzschild radius $R_s = 2M G_N/c^2 $. If $L\le R_s$, the rod is not a rod anymore, rather, it will look like a
 \emph{black hole!} This is the so-called ``hoop-conjecture'': any physical object extending along a certain direction less 
than its Schwarzschild radius, \emph{collapses into a black hole}  \cite{thorne}. How a black hole appears in a boosted frame
is an overlooked problem except in the somewhat ambiguous ``~shock wave limit~'' where $\gamma\to \infty$, $M\to 0 $ while the 
product is kept finite, i.e. $0< \gamma M < \infty $ \cite{Aichelburg:1970dh}.
\end{itemize}

By considering both arguments at the same time, one expects that Quantum Gravity 
imposes intrinsic limits to the relativistic  contraction of physical objects. Presently, the most promising candidate for
a self-consistent theory of quantum gravitational phenomena is Super-String Theory. From its vantage point, String Theory
``solves'' the problem from the very beginning by assuming that the building blocks of everything are finite length, vibrating 
strings. Nothing can be ``smaller,'' in the sense that  any distance (length) smaller than the string length 
$\sqrt{\alpha^\prime}$ does not have physical meaning. As string theory is a quantum theory of gravity,  the string
length may be identified with the Planck Length $L_P \approx 10^{-33}\, cm $. Unfortunately, to our knowledge string theory 
says nothing about the Lorentz-Fitzgerald contraction and how to modify it.\\
Our foregoing discussion, on the other hand, requires that
any quantum gravity-inspired extension ought to contain both Newton and Planck constant, $G_N$ and $\hbar$, and 
reproduce (\ref{contract}) when $G_N$ or $\hbar$ are ``switched-off''.\\
According to the hoop-conjecture there must be a \emph{critical boost} factor $\gamma_\ast\equiv 1/ \sqrt{1-\beta^2_\ast} $ 
that characterizes the transition from a gravitationally interacting two-particle system into  a black hole.
In order to determine $\gamma_\ast$, let us tentatively change the
contraction formula into the following expression,

\begin{equation}
\tilde{ L}\left(\, \beta\,\right) = L_0\, \sqrt{1-\beta^2} 
+\frac{L_\ast^2\, \theta_H\left(\, \beta\,\right)}{4L_0\, \sqrt{1-\beta^2}}
\label{ltilda}
\end{equation}
where, $ \theta_H $, is the Heaviside step function which guarantees that the extra term does not affect the measure of 
$ L $ at rest \footnote{We define $\theta_H(x)$ as
\begin{eqnarray}
 \theta_H(x)=1 \longleftarrow x>0\ , \nonumber\\
 \theta_H(x)=0 \longleftarrow x\le 0\ , \nonumber
\end{eqnarray}
Sometimes, it is conventionally chosen  $\theta_H(0)\equiv 1/2$.  In this case, a $\beta$-independent 
quantity $L_\ast^2/8L_0$ must be subtracted in (\ref{ltilda})}. 
Moreover, since any macroscopic length  is tens of orders of magnitude larger than $L_\ast $, the second term in (\ref{ltilda})
gives a relevant contribution only in the ultra-relativistic regime $\beta\approx  1$.
The minimum of the function $\tilde{ L}\left(\, \beta\,\right) $ is

\begin{equation}
\frac{d\tilde{ L}}{d\beta} =0  \longrightarrow \gamma_\ast = \frac{2L_0}{L_\ast}\ , 
\quad \tilde{ L}\left(\, \beta_\ast\,\right) =L_\ast
\label{lmin}
\end{equation} 

For $\gamma > \gamma_\ast$ the function $ \tilde{ L}\left(\, \beta\,\right) $ ``~bounces back~'' and increases as stipulated 
in our earlier discussion on the basis that  
a similar behavior is shown by a fundamental string which cannot shrink below its minimal length $l_s=\sqrt{\alpha^\prime}$ 
while increasing its energy excites  
 higher and higher vibration modes forcing the string to elongate. Thus, a natural 
choice for $L_\ast $ is  $L_\ast=l_s=\sqrt{\alpha^\prime}$. For later convenience, it seems also worth recalling again that
a highly excited  string looks rather like a black hole. With this identification, equation(\ref{lmin}) tells that in any 
inertial reference frame no physical length can be smaller that the string length:

\begin{equation}
\tilde{ L}\left(\, \beta\,\right)\ge \sqrt{\alpha^\prime}
\end{equation}

and the critical boost representing the turning point between contraction and dilatation  
turns out to be $\gamma_\ast =2L_0/\sqrt{\alpha^\prime}$.  \\
Now, let us take a closer look at equation (\ref{ltilda}) by way of some illustrative examples: 
\begin{itemize}
\item
Take for $L_0$ the Compton wavelength $\lambda_C=  1/m $ of a particle.  
In analogy with the string improved GUP, we obtain the following modified de Broglie formula

 \begin{equation}
\tilde{ \lambda}\left(\, \beta\,\right) = \lambda_C\, \sqrt{1-\beta^2} 
+\frac{\alpha^\prime\,\theta_H\left(\, \beta\,\right) }{4\lambda_C\, \sqrt{1-\beta^2}}
\label{compt}
\end{equation}

As $ \lambda \left(\, \beta\,\right)$ cannot be smaller than $\sqrt{\alpha^\prime}$ it follows that the mass spectrum of
an ``~elementary~'' particle is bounded from above by the limiting mass $1/\sqrt{\alpha^\prime}$

\begin{equation}
\lambda \ge \sqrt{\alpha^\prime}\longrightarrow m\le \frac{1}{\sqrt{\alpha^\prime}}
\end{equation}
 \item
Next, consider the case of a Schwarzschild black hole, $L_0=R_s$. Equation (\ref{ltilda}) tells us how the horizon
radius will appear from a Lorentz boosted frame

 \begin{equation}
R_s\left(\, \beta\,\right) = 2MG_N\, \sqrt{1-\beta^2} +\frac{\alpha^\prime\theta_H\left(\, \beta\,\right)}{8MG_N\, 
\sqrt{1-\beta^2}}
\label{rsboosted}
\end{equation}
The first term shows how the Schwarzschild radius of a moving mass appears contracted as any other physical length.
The second term in (\ref{rsboosted}) takes into account the existence of a \emph{``hard core'' characterized by a universal, 
unsurpassable linear energy density, or tension, that prevents further contraction, or collapse into a point singularity.} We shall come back to this essential point in the concluding section of this paper.\\
The critical boost is 

\begin{equation}
\gamma_\ast =\frac{4MG_N}{\sqrt{\alpha^\prime}}
\end{equation}

 For $\gamma=\gamma_\ast $ the Schwarzschild radius reaches its minimal value 
$R_H\left(\,\beta^\ast\,\right)=\sqrt{\alpha^\prime}$. A snapshot of a black hole at this \emph{minimal size} will
show an object with an \emph{effective mass} $M_\ast$ defined as

\begin{equation}
R_H\left(\,\beta^\ast\,\right)\equiv 2M_\ast G_N\longrightarrow M_\ast=\frac{\sqrt{\alpha^\prime}}{2G_N}
\end{equation}

In a string theoretical formulation of quantum gravity, the Regge slope can be related to  the Newton constant through
$\alpha^\prime =2G_N $. Thus, we find $M_\ast = 1/\sqrt{2G_N}=M_P$ and $ R_H\left(\,\beta^\ast\,\right)=L_P$.\\
If we formally assign to the black hole a Compton wavelength $\lambda_{BH}\equiv 1/ M $, we can write equation (\ref{rsboosted})
as follows,
\begin{equation}
R_s\left(\, \beta\,\right) = \frac{L_P^2}{\lambda_{BH}}\, \sqrt{1-\beta^2} 
+\frac{\lambda_{BH}\, \theta_H\left(\, \beta\,\right)}{4\, \sqrt{1-\beta^2}}
\label{rsboosted2}
\end{equation}

Comparison with equation (\ref{compt}) shows that $\lambda_{BH}$ enters the modified contraction law in the inverse
way with respect to $\lambda_C$, thus suggesting that the de Broglie wavelength of a black hole can be written as

\begin{equation}
\lambda^{dB}_{BH}\equiv \frac{\lambda_{BH}}{\sqrt{1-\beta^2}}=\gamma\, \lambda_{BH}
\end{equation}

Once the critical boost $\gamma_\ast = 2M\sqrt{\alpha^\prime}=M/M_P$ is passed, the first term in (\ref{rsboosted2}) 
is negligible and the Schwarzschild radius expands:

\begin{equation}
R_s\left(\, \beta\,\right) \approx \frac{\lambda^{dB}_{BH}}{4}.
\label{rsultra}
\end{equation}

It is worth mentioning that, recently, a new family of singularity-free  black hole-metrics was reported in 
\cite{Nicolini06,Ansoldi:2006vg,Spallucci:2008ez,Nicolini:2008aj,Smailagic:2010nv,Nicolini:2009gw}, where the 
existence of a minimal length is assumed at the outset in the Einstein equations.  A remarkable properties of these black holes
is to admit extremal configurations even in the neutral non-spinning case. Extremality corresponds to
the lowest mass state of the system and to a minimal radius of the event horizon which equals  few times the
minimal length.  In some simple models, it is possible to choose the free length scale that regularizes the short distance 
behavior in such a way that the radius of the extremal configuration is exactly the Planck 
length \cite{Spallucci:2011rn,Spallucci:2012xi,Nicolini:2012fy}. Without  introducing
the improved Lorentz law, the very idea of a minimal size object would become observer-dependent. 
\end{itemize}

To sum up, at this point we have :
\begin{enumerate}
\item
equation (\ref{ltilda}) for a (semi)classical length $L_0$ with  string corrections;
\item equation (\ref{compt}) for the deBroglie wave length with string correctins;
\item equation (\ref{rsboosted}) for the Schwarzschild radius with string corrections.
\end{enumerate}

Now, it is time to consider the \emph{hoop conjecture} and check the self-consistency of our formulae.\\
Let us start with case (1) and address the central question:  \emph{Can a boosted object be seen contracted below its 
Schwarzschild radius?} \\
If so, the hoop conjecture would imply the original object is seen as a black hole... ?

\begin{equation}
L_0\, \sqrt{1-\beta^2} +\frac{\alpha^\prime\theta_H\left(\, \beta\,\right)}{4L_0\, \sqrt{1-\beta^2}}\le R_s\,
 \sqrt{1-\beta^2} +\frac{\alpha^\prime\theta_H\left(\, \beta\,\right)}{4R_s\, \sqrt{1-\beta^2}}
\label{hoopconj}
\end{equation}

We may regard this relation either as an equation for the radius $L_0$ below which the object is shielded by an horizon, 
or as an equation for a hypothetical ``terminal speed'' $\tilde\beta$ that, once surpassed, will make the object to appear 
inside its own Schwarzschild hoop. In the first
case, it is immediate to recognize that $L_0\le R_s$ is the $\beta$-independent, somewhat ``~trivial~'' solution. 
In order to be seen as a black object, the maximal length, at rest, must be smaller than the Schwarzschild radius $R_s$.  
On the other hand, if one assumes that 
$L_0>R_s$ and tries to determine $\tilde\beta$ from (\ref{hoopconj}), then one finds: 

\begin{equation}
\tilde\beta^2 = 1 + \frac{\alpha^\prime}{4R_s\, L_0} >1
\end{equation}
which is \emph{unphysical}. Only by moving at a speed greater than the speed of light can an object turn into a black hole.
Thus, \emph{even in the presence of quantum corrections, there is no inertial frame where a classical object with linear
size $L_0> R_s$ may appear as a black hole.}\\
Let us consider now a quantum particle, rather than a classical object.

\begin{equation}
\lambda_C\, \sqrt{1-\beta^2} +\frac{\alpha^\prime\theta_H\left(\, \beta\,\right)}{4\lambda_C\, \sqrt{1-\beta^2}}\le 
R_s\, \sqrt{1-\beta^2} +\frac{\alpha^\prime\theta_H\left(\, \beta\,\right)}{4R_s\, \sqrt{1-\beta^2}}
\end{equation}

Once more, the ``~terminal boost~'' is unphysical, i.e. $\beta >1$, and the only acceptable solution is 

\begin{equation}
\lambda_C=R_s \longrightarrow m=\frac{1}{\alpha^\prime}=M_P
\end{equation}

Thus, as before, there does not exist an inertial frame where an \emph{isolated} elementary particle with (invariant) 
mass $m< M_P$ looks like a black hole. However, this
conclusion does not prevent the production of a black hole in the final state of a two-body high energy scattering where 
the hoop conjecture has been validated using numerical/computational techniques
\cite{Choptuik:2009ww}. This  different case will be discussed in the next section using a more analytical approach.

\subsection{High energy collisions and black hole production}

We are now ready to extend (\ref{compt}) to the case of a two-body system of colliding partons
in the framework of higher dimensional quantum gravity. In this case, the gravitational coupling
constant is $G_\ast $ with dimensions (in natural units) $\left[\,G_\ast\,\right]= l^{d-1} $, much
below the Planck energy, and $d$ is the number of space-like dimensions ($\ge 3$ ).  If the two partons
have four-momenta $p_1$ and $p_2$, it is useful to introduce the Mandelstam variable $s=-\left(\, p_1 + p_2\,\right)^2$.
In terms of $s$ we can define the ``\emph{effective Schwarzschild radius}'' of the system as

\begin{equation}
 r_H\left(\, s\,\right) = \left(\, 2\sqrt{s} \, G_\ast \,\right)^{1/(d-2)}\equiv 
\left(\, \sqrt{s} \, L_\ast^{d-1} \,\right)^{1/(d-2)}
\end{equation} 
 
where $L_\ast $ is the higher dimensional minimal length. 
The hoop-conjecture states that whenever the two partons collide with an \emph{impact parameter} 
$b\le r_H\left(\, s\,\right) $, then a micro-black hole is produced. In our approach we can rephrase this
statement as follows: the two-parton system will collapse into a black hole if the de Broglie wavelength
(\ref{compt}) is smaller, or equal, to the Schwarzschild radius (\ref{rsboosted})

 \begin{equation}
\frac{1 }{\sqrt{s}}     + \frac{L_\ast^2}{4}\sqrt{s} \le 
\left(\, \sqrt{s}\, L_\ast^{d-1} \,\right)^{1/(d-2)} 
+ \frac{L_\ast^2}{4}\left(\, \sqrt{s}\, L_\ast^{d-1} \,\right)^{-1/(d-2)}
\label{lambdeff}
\end{equation} 

where we have switched to natural units, $\hbar=c=1$ and no step-function is needed as the two particle are \emph{by definition}
in a relative state of motion. Solving for $s$ we find the \emph{threshold} invariant
energy for the creation of a micro black hole. This is a \emph{necessary, but not sufficient} condition for
this event to occur. As it can be expected, the production channel
opens up once the quantum gravity energy scale is reached

\begin{equation}
 \sqrt{s}\ge \frac{1}{L_\ast}= M_\ast
\label{lambdalast}
 \end{equation}

Equation (\ref{lambdalast}) tells us that in a high energy scattering experiment we can probe
distances down to $L_\ast$ but not beyond. \emph{The would be trans-Planckian region is shielded
by the creation of a black hole with linear dimension increasing with $s$.} This argument
is the essence of a recent proposal by Dvali end co-workers \cite{Dvali:2010bf,Dvali:2010ue}
to explain how quantum gravity can self-regularize in the deep ultraviolet region \cite{Spallucci:2011rn}
\footnote{The scenario of UV self-complete quantum gravity is especially attractive
when realized in the more general framework of $TeV$ quantum gravity. In this case, the Planck scale is lowered

 down to the $TeV$ scale and opens the exciting possibility to detect quantum gravity signals at LHC.}. 
Thus, \emph{the trans-Planckian regime is actually inaccessible, and the deep UV region is dominated by large, ``classical''
field configurations.} This mechanism has been dubbed ``classicalization.'' 
\cite{Dvali:2010jz,Dvali:2010ns}.\\
There is a second important consequence of the relation (\ref{lambdeff}) regarding the final
stage of \emph{black hole evaporation.} Micro black holes are known to be semi-classically unstable
because of Hawking radiation. However,  the standard description of thermal decay breaks 
down when the black hole approaches the full quantum gravity regime. Even worse, no
semi-classical model can foresee the end-point of the process which remains open to largely unsubstantiated speculations. \\
Equation (\ref{lambdeff}) on the other hand, shows not only the transition of a two-particle system $\rightarrow$ black hole, 
but the inverse process as well.  Start from the black hole region and decrease the invariant mass
of the object. Effective models of ``~quantum gravity-improved~'' black holes suggest that
for $ M>> M_\ast $ the semi-classical model is correct and the particle emission is to a good
degree of approximation a grey-body thermal radiation.  However, as $M \to M_\ast $ and the
size of the black hole becomes comparable with $ L_\ast $, the mass of the object reveals a
discrete spectrum and the decay process goes on  through the emission of few quanta while
jumping quantum mechanically towards the ground state. In this late stage of decay the
black hole behaves like a hadronic resonance, or an unstable nucleus,  rather than a hot body.
Thus, it is not surprising that after crossing the critical point $M=M_\ast $ one is left with an
``~ordinary~'' elementary particle system \cite{Meade:2007sz}.

\section{Final remarks: a new ``~black hole universal constant~'' and the existence of a maximal force in Nature}

In this note we have proposed a consistent framework to reconcile the existence of a new fundamental
constant of Nature, with length dimension, with the Lorentz-Fitzgerald contraction expected from
Special Relativity. The presence of an ultimate length barrier has been related to the presence
of a black hole barrier that shields the trans-Planckian regime from a direct investigation.  
The critical boost factor $\gamma_\ast$ that marks the sharp transition 
from Special Relativity to the ``quantum gravity'' regime has been related to the threshold energy where
gravitationally interacting point-particles collapse into an extended micro black hole. This threshold energy
is determined by the final unification scale where quantum gravity becomes
as strong as the other interactions. \\
\emph{Assuming that the ``~super-unification~'' scale is the Planck scale, is there any clue as to what the expression of 
the ``~super-unified force~'' might be?} This question leads us to confront the notion of  \emph{maximal tension} 
introduced by Gibbons' in the framework of General Relativity \cite{Gibbons:2002iv}.\\
Gibbons' conjecture is that there exists in nature a 
limiting force, let's call it a \emph{superforce}, \footnote{The idea of a maximal force, naturally leads 
to the idea of ``\emph{maximal accelration}''. A similar suggestion has been recently presented in the framework of loop quantum gravity 
\cite{Rovelli:2013osa}. } 
whose \emph{exact expression} is given by: 

\begin{equation}
F_s = \frac{c^4}{4G_N}\label{gp}
\end{equation}

With the above expression in hands, we are in a position to add some final remarks that may shed a different light
on the whole sequence of arguments presented in this paper. A more comprehensive account of the following points will 
be presented in a forthcoming article \cite{noi2}.

\begin{enumerate}
\item Note, first, that our definitions of Planck units are consistent with
Gibbons' expression of the super-force. In other words, on \emph{dimensional grounds} alone, the superforce is the ``~Planck 
force.~'' Having established that, it takes an elementary calculation to verify that the Gibbons-Planck force 
$ F_s=c^4/4G_N $ is, indeed, the Planckian limit of both the electrostatic Coulomb force and the static gravitational 
Newton's force! While this is a definite clue that (\ref{gp}) is the unification point of the electro-gravitational force, 
it remains an open question whether it also represents the super-unified value of all fundamental forces.
\item With hindsight, the conspicuous absence of $\hbar$ from the Gibbons-Planck expression seems to 
support the \emph{classicalization} idea as well as the idea of a transition from ``~contraction~'' 
to ``~dilation~'' in the modified expression of the Lorentz-Fitzgerald formula. Again, with hindsight, both ideas are inherent in 
the fundamental relationship (\ref{intersect}):

\begin{equation}
\frac{\hbar}{M_\ast c} = \frac{2 M_\ast G_\ast}{c^2}
\end{equation}

As a matter of fact, inspection of the above equation shows that, on the one hand it defines the Planck scale of mass-energy, 
but, on the other hand, it signals a trade-off, \emph{at the Planck scale of energy,} between a quantum length (Compton) and 
a classical one (Schwarzschild) with the concomitant transition from ``~Lorentz contraction~'' in the particle phase 
to ``~Schwarzschild expansion~'' in the black hole phase.
\item  The appearance of $ G_N $ 
in $F_s$ makes one wonder about the specific role of gravity in the unification of fundamental forces. 
Here we offer an alternative, gravity inspired, interpretation of the superforce: it represents the ultimate \textit{linear} 
energy density of a black hole. In order to underscore this point, consider the conventional
volume density of a body. In the case of a black hole this leads to the somewhat counterintuitive result that the density is 
inversely proportional to the square  of the mass 

\begin{equation}
\rho_{BH} = \frac{M}{4\pi R_s^3/3} =\frac{3c^6}{32\pi G_N^3}\frac{1}{M^2}
\end{equation}

so that, while mini black holes may possess a nuclear density, galactic black
holes can be less dense than water. On the other hand, by considering the
\textit{linear} energy density, one obtains the \emph{universal constant}

\begin{equation}
\rho_\ast = \frac{M c^2}{2 R_s} =\frac{c^4}{4 G_N}
\label{super}
\end{equation}

In words, there exists in Nature a limiting linear density that is 
a \textit{universal} characteristic of all (Schwarzschild) black holes \textit{regardless
of their mass or size}. At first sight this result may seem surprising and
hard to understand. In actual fact, the physical explanation
rests on the \textit{duality} between deep UV and far IR domains in quantum gravity.
The unique properties of black holes bridge the gap between trans-Planckian
and classical physics \cite{Kosyakov:2007zv}! \\
\item Last, but not least, given the background of ideas advanced in this article, it seems natural to identify $\rho_\ast $ 
with the energy density of a \textit{relativistic string} and, therefore, 
we identify the  super-force Eq.(\ref{super}) with the universal \textit{string tension}

\begin{equation}
\rho_\ast\equiv \frac{\hbar c}{2\pi\alpha^\prime}\equiv T_s
\label{superstring}
\end{equation}

Therefore, there are two equivalent ways of writing $\rho_\ast$:

\begin{itemize}
\item
classical, \emph{macroscopic} form given by Gibbons' \emph{ Maximal Tension} (\ref{super});
 \item
quantum, \emph{microscopic} form which is the \emph{String Tension} (\ref{superstring})
\end{itemize}

The two definitions are linked  through:\\
i) the existence of a universal linear energy density for black holes exposing their ``stringy nature''.
\cite{Horowitz:1996nw};\\
ii) the ``classicalization mechanism'' of quantum gravity that identifies trans-Planckian black holes with classical,
macroscopic objects.  \\
\emph{It seems to be a unique property of gravity to bridge the gap between micro and macro worlds.}
\end{enumerate}

\end{document}